\documentclass[%
 aip,
 amsmath,amssymb,
 reprint,%
]{revtex4-1}

\usepackage{graphicx}
\usepackage{dcolumn}
\usepackage{bm}

\usepackage[utf8]{inputenc}
\usepackage[T1]{fontenc}
\usepackage{mathptmx}

\begin{document}

\preprint{AIP/123-QED}

\title[Robustness of the mental lexicon]{Multiplex networks quantify  robustness of the mental lexicon to catastrophic concept failures, aphasic degradation and ageing}

\author{Massimo Stella}
\email{massimo.stella@inbox.com}
\affiliation{ 
Complex Science Consulting, Via Amilcare Foscarini 2, 73100 Lecce, Italy
}%

\date{\today}

\begin{abstract}

Concepts and their mental associations influence how language is processed and used. Networks represent powerful models for exploring such cognitive system, known as mental lexicon. This study investigates lexicon robustness to progressive word failure with multiplex network attacks. The average lexicon of an adult English speaker is built by considering 16000 words connected through semantic free associations and phonological sound similarities. Progressive structural degradation is modelled as random and targeted attacks. Words with higher psycholinguistic features (e.g. frequency, length, age of acquisition, polysemy) or network centrality (e.g. closeness, PageRank, betweenness and degree) are targeted first. Aphasia-inspired attacks are introduced here and target first words named correctly, more or less frequently, by patients with anomic aphasia, a pathology disrupting word finding. Robustness is measured as connectedness, fundamental for activation spreading and lexical retrieval, and viability, a multi-layer connectivity identifying language kernels. The lexicon is resilient to random, aphasia-inspired and psycholinguistic attacks. Catastrophic phase transitions happen when phonological and semantic degrees are combined, making the lexicon fragile to multidegree attacks. The viable kernel is fragile to multi-PageRank and to aphasia-inspired attacks. Consequently, connectedness in the lexicon is mediated by hubs, whereas viability enables a lexical semantic/phonological interplay and corresponds to a facilitative naming effect in aphasia. These effects persist also through ageing, in different network representations of younger and older lexicons. This study indicates the need to prevent failure of high multidegree and viable words in the mental lexicon when pursuing the design of effective language restoration strategies against cognitive impairing.

\end{abstract}

\maketitle

\section{Introduction}

Complex networks constitute a powerful tool \cite{baronchelli2013networks,siew2018cognitive} for investigating how conceptual associations, as represented in the human mind, influence cognitive computing for language acquisition and use  \cite{aitchison2012words,vitevitch2018spoken,doczi2019overview}. This \textit{mental lexicon} of concepts and their associations has been successfully investigated by several small-scale network models in psycholinguistics \cite{tiptongue1966,dell1997lexical,aitchison2012words,erdeljac2008syntactic,doczi2019overview}. The recent advent of network science \cite{newman2018networks} further promoted the development of large-scale network models, enabling new understanding of how large ensembles of concepts are structured and influence language processing in the semantic/meaning-driven\cite{de2016large,de2013better,beckage2016language,kenett2011global,zemla2019analyzing} and phonological/sound-driven \cite{vitevitch2008can,stella2015patterns,siew2016spokenrecall,vitevitch2018spoken,neergaard2019constructing} aspects of the lexicon. One of the strongest appeal of such network models is to build a given layout of conceptual associations and then observe how it degrades when some damage is done \cite{borge2011modeling,borge2012topological}, e.g. some links or nodes are made unavailable to the system. These quantitative experiments are called \textit{network attacks} and are used for investigating the robustness of networked systems to random or targeted damage \cite{callaway2000network,frantz2009robustness,iyer2013attackrobustness,zhao2016robustness,newman2018networks,baxter2018targeted}. Although it is appealing to investigate robustness of single aspects of the mental lexicon to degradation, this approach would neglect the extensive evidence indicating how failed lexical retrieval \textit{crucially} depends on the interplay between semantic and phonological conceptual similarities   \cite{tiptongue1966,aitchison2012words,erdeljac2008syntactic,stella2018multiplex,castro2019multiplex}. In order to embrace multiple aspects of the mental lexicon in a single network representation, this work suggests studying lexicon robustness through a minimal representation of semantic and phonological associations for 16000 English words as a multiplex lexical network \cite{stella2017multiplex,stella2018multiplex,stella2018cohort,stella2019viability,stella2019modelling}. Such representation is used as a quantitative framework for exploring the large-scale robustness of the mental lexicon under progressive word failure through the formalism of network-based attacks \cite{frantz2009robustness,borge2012topological,iyer2013attackrobustness,baxter2018targeted}. Finding more or less efficient ways of disrupting connectivity within words in the mental lexicon can offer valuable insights about the cognitive organisation of words in the mind and, more importantly, also provide useful information aimed at understanding and then hampering or preventing cognitive decline in clinical pathologies where language is progressively disrupted, like aphasia \cite{roach1996philadelphia,erdeljac2008syntactic} or Alzheimer's disease \cite{zemla2019analyzing,zemla2018estimating}.

Contrary to physical systems, that can be reproduced and tested directly in the laboratory, the mental lexicon remains a cognitive construct, i.e. a system that cannot be accessed or tested straight way \cite{aitchison2012words}. Once hypothesised from the structure of language, conceptual associations in the mental lexicon can only be indirectly tested through psycholinguistic tasks, whose outcomes can provide indirect evidence about the importance of the original associations for language processing and acquisition \cite{vitevitch2018spoken}. In the last forty years, overwhelming empirical evidence at the interface of computer science and psycholinguistics has reported the relevance of semantic relationships such as free associations \cite{kenett2014investigating,kenett2017semantic,de2018small} or sound similarities like phonological edit distances \cite{vitevitch2008can,stella2015patterns,neergaard2019constructing} for predicting and understanding a variety of processes like lexical retrieval in recall experiments and identification responsiveness \cite{vitevitch2018spoken,de2013better}, language acquisition \cite{beckage2016language,siew2018cognitive}, writing styles \cite{amancio2012using}, knowledge structuring and search \cite{kenett2017semantic,goldstein2017influence} and even creativity levels \cite{kenett2014investigating,kenett2019can}. All these approaches underline the importance of semantic and phonological associations between concepts for better understanding how language is perceived and processed in the human mind. Single-layer complex networks have been a valuable asset in opening the way to large-scale, quantitative investigations about the global layout and organisation of the mental lexicon \cite{baronchelli2013networks,beckage2016language,siew2018cognitive}. Recently, thanks to advancements in the field of multiplex networks \cite{battiston2017new,bianconi2018multilayer}, a novel generation of large-scale models known as \textit{multiplex lexical networks} and aimed at achieving multiplex lexical representations have been suggested and successfully tested for better understanding phenomena like language processing \cite{stella2018cohort}, creativity and fluid intelligence \cite{stella2019viability} and early word acquisition \cite{stella2017multiplex,stella2019modelling}.

This work adopts a large-scale multiplex representation of words in the mental lexicon by means of the two most powerful single-layer networks used in previous studies, free associations and phonological similarities, with the aim of studying how combining semantics and phonology in the mental lexicon affects robustness to progressive word failure, a phenomenon where words and their associations become increasingly inactive \cite{borge2012topological}. 

By drawing inspiration from statistical physics and network science \cite{frantz2009robustness,iyer2013attackrobustness}, the robustness of the mental lexicon is tested by both random attacks (where nodes are made inactive uniformly at random) and targeted attacks (where most prominent/central nodes are made inactive first as indicated by a certain attack strategy). Robustness is quantified as indicated in previous approaches \cite{frantz2009robustness,arbesman2010structure} to network robustness under node removal, which is in terms of connectedness. Given the multiplex nature of the current system, connectedness has to be clearly defined in this approach. In agreement with previous literature \cite{bianconi2018multilayer}, connectedness here is tested by the existence of any sequence or path of semantic and/or phonological associations connecting words. Hence, connectedness on the multiplex network assumes the possibility to freely transition between semantic and phonological aspects of the mental lexicon. This minimal assumption, while simplistic, has the advantage of avoiding the quantification, if any, of a cognitive cost \cite{bianconi2018multilayer} for transitioning between semantics and phonology, which is also an open question for the relevant psycholinguistic literature \cite{aitchison2012words} and for most empirical multiplex networks \cite{battiston2017new,bianconi2018multilayer}. Importantly, the combination of semantic and phonological layers of conceptual knowledge opens new ways for connecting concepts that could be absent in the starting individual networks when kept as separate. An example is the work by Stella \cite{stella2018cohort}, where the multiplex combination of semantic or phonological layers highlighted connectivity patterns related to phonological priming\cite{aitchison2012words,vitevitch2018spoken} and absent in the individually considered networks. The size of the largest connected component of the attacked multiplex network is not only a commonly used metric of robustness in network science \cite{frantz2009robustness,iyer2013attackrobustness} but it also has a clear cognitive relevance for lexical processing. According to the revised spreading activation model by Bock and Levelt \cite{bock1994language}, a powerful model of psycholinguistics for explaining positive semantic and phonological priming in lexical retrieval, words are identified and recalled through an \textit{activation signal} that propagates over the mental lexicon along conceptual associations encapsulating both semantic and phonological aspects of words. While decaying over time, activation can bounce back between conceptually associated words and also concentrate on the desired concept, that can be subsequently identified and recalled. Activation cannot spread outside of conceptual connections. While inhibitory connections can also be present \cite{bock1994language,baronchelli2013networks,doczi2019overview}, connectedness represents a first, fundamental constraint driving lexical processing. For this reason, this study will adopt multiplex connectedness as a metric for robustness of the mental lexicon under attack.

Notice that the above notion of connectedness is not the only one possible in a multiplex network. Rather than using paths of conceptual associations being made of either semantic or phonological network, a more restrictive "AND" logic could be used, instead. Baxter and colleagues \cite{baxter2016unified} identified \textit{viable connectivity} in terms of any two nodes being connected within each and every layer, e.g. paths connecting the same two words through free associations only and, at the same time, through phonological associations only. The largest viable component of the mental lexicon was identified \cite{stella2018multiplex} as a network core for the mental lexicon, thus facilitating network navigation, and emerging through an explosive phase transition in children around age 7-8 yrs. Subsequent work \cite{stella2019viability} identified the largest viable cluster as a relevant component of the mental lexicon, made of more frequent, easier to identify, earlier acquired words, of relevance also for knowledge exploration across groups with different levels of creativity. Building upon such evidence for the cognitive relevance of the largest viable component in the mental lexicon, also its size will be adopted as a metric for robustness under attack.

In addition to attacking the mental lexicon through psycholinguistics-driven strategies (e.g. attack first words with more meanings) and network-driven strategies (e.g. attack first words of higher closeness centrality), this work tests also strategies based on clinical data from the Philadelphia Naming Test \cite{roach1996philadelphia} (PNT). By training a logistic regression model over a dataset of picture naming by people with \textit{anomic aphasia}, the probability for correct naming of a given word in the whole multiplex lexical network was predicted based on psycholinguistic and network metrics of relevance, reminiscently of previous network approaches in people with aphasia \cite{castro2015using,castro2019multiplex}. The training dataset included 31000 picture naming attempts of 173 words. Although relatively small, the adopted dataset represented valuable knowledge for approximating the difficulty people with aphasia have in retrieving and producing a given word. Testing also aphasia-like attack strategies based on these picture naming probabilities represented an interesting large-scale and immediate test for identifying how the connectivity in the mental lexicon degrades and potentially design training approaches aimed at hampering such decline.

Identifying the robustness of the mental lexicon to all the above attacks can open new ways for understanding, within the due assumptions and approximations, how language declines in specific pathologies and therefore come up with efficient countermeasures \cite{kirrie2018}. Notice that the main assumption of this analysis is that words fail progressively and the corroboration of such working hypothesis opens new challenges for clinical fieldwork. In addition to the clinical implications, robustness results represent also a valuable source of information \cite{borge2012topological} for exploring and understanding how the human mental lexicon is structured and behaves under degradation.

This study is structured as follows. The Methods section reports on the data and network metrics used in the attack experiments. The Results section is divided in three parts: The first one outlines the robustness profile of the mental lexicon in terms of global connectedness; the second part focuses on the viable component; the third part tests lexicon robustness through ageing when semantic connections get weakened. A final Discussion compares and interprets results and identifies concrete future research directions opened by the current approach.

\section{Methods}

\subsection{Cognitive data}

Four psycholinguistic features of words were used in this study: (i) frequency of words in language as computed in SUBTLEX \cite{brysbaert2009moving} (a dataset including 2.2 billions subtitles), (ii) word length, (iii) mean age of acquisition of a concept in English native speakers \cite{brysbaert2017test} and (iv) number of meanings of a word (polysemy score) computed from the dictionary implemented in WordData[] by WolframResearch and based on WordNet 3.0 \cite{miller1995wordnet}. All these measures were reported as being predictive of a variety of language processing tasks such as lexical identification \cite{brysbaert2009moving,aitchison2012words,brysbaert2017test}, recall \cite{brysbaert2009moving,siew2016spokenrecall} and confusability \cite{brysbaert2009moving,vitevitch2018spoken,brysbaert2017test}. Based on such evidence, the above psycholinguistic features were used here as indicators for designing targeted attack strategies to the multiplex lexical network of relevance from a cognitive perspective.

\subsection{A logistic regression model for estimating correct naming in people with anomic aphasia}

With the aim of having a data-driven way of simulating progressive decline of the mental lexicon as affected by a clinical pathology, this study used cognitive data from the Philadelphia Naming Test (PNT) \cite{roach1996philadelphia}. The PNT dataset, available through a free registration on the Moss Aphasia Psycholinguistics Project Database website, reports on the performance of people with aphasia and healthy controls in a naming task including 175 items/words \cite{mirman2010large}. People with a diagnosis for aphasia were shown a picture and asked to name it. Each picture represented a concept, which was correctly named by almost all healthy subjects (with a probability above 0.99). People with aphasia had degraded linguistic skills, so that to them picture naming was more difficult \cite{martin1998lexical}. 

Considering almost 31600 utterances produced by people with \textit{anomic aphasia}, the probability for correct picture naming in people with aphasia was 0.806. In general, people with anomic aphasia are characterised by an overall fluent speech complicated by occasional difficulties in word finding, happening with higher frequency in comparison to healthy subjects \cite{martin1998lexical,hillis2007aphasia,laine2013anomia}.

By aggregating together all naming experiments relative to each single item/word, empirical probabilities $p_w$ for correct naming were obtained for all the 173 words from PNT present also in the multiplex lexical network. These probabilities were then fed to a logistic regression model based on predictors available in the multiplex network, namely the above mentioned psycholinguistic features and also network metrics \cite{newman2018networks} like single-layer and multiplex degree, PageRank, closeness and betweenness centrality. The selection of these network metrics was based on previous results from \cite{castro2019multiplex}, which used a similar approach for predicting correct naming in people with aphasia. The regression analysis was fitted through a multi-step procedure, as implemented in R, in order to discard redundant predictor variables and minimises Akaike's Information Criterion. 

\begin{table}
\caption{\label{tab:statistics}Results of the logistic regression of empirical production probabilities $p_w$ by using psycholinguistic and network predictors. The more stars, the lower the p-value. The multi-step procedure discarded network metrics and identified the model minimising Akaike's Information Criterion (28709) in 5 Fisher steps.}
\begin{ruledtabular}
\begin{tabular}{ccccc}
Coefficient & Estim. & St.Err. & z-score & p-value \\
\hline
(Intercept) & -0.466 &   0.120 & -3.87 & 0.0001 ***\\
Freq. & -0.073 &   0.022 & -3.28 & 0.0010 ** \\
Length  &        -0.326  &  0.026 & -12.55 & $< 10^{-8}$ *** \\
Age of Acq.  &        -0.721  &  0.037 & -19.22 & $< 10^{-7}$ *** \\
Polysemy    &       0.069  &  0.017 &  3.91 & $< 10^{-5}$ *** \\
Asso. Clos.    &       1.100 &   0.198 &  5.53 & $< 10^{-7}$ *** \\
Mult. Clos.      &    -0.221  &  0.060 & -3.68 & 0.0002 *** \\
Asso. PageR.    &      0.709  &  0.142 &  4.97 & $< 10^{-7}$ *** \\
Phon. PageR.     &      0.059  &  0.023 &  2.57 & 0.0101 *  \\
Asso. Degree    &     -0.492  &  0.141 & -3.47 & 0.0005 *** \\
Asso. Betw. &     -0.239  &  0.038 & -6.17 & $< 10^{-7}$ ***\\
\end{tabular}
\end{ruledtabular}
\end{table}

The fitted logistic model reported a pseudo-$R^2$ of 0.096 (Cragg and Uhler's). All psycholinguistic variables were identified as having an influence over correct naming predictions. Multiplex closeness was the only multiplex measure to display predictive power over correct naming performance. For degree, PageRank and betweenness, single-layer measures were selected to be within the model. Multiplex degree indicate the sum of degrees across layers \cite{battiston2017new}. Multiplex PageRank was based on versatility \cite{de2015ranking,bianconi2018multilayer}. Multiplex betweenness and centrality exploited multilayer paths, like in previous works \cite{stella2018cohort,stella2018distance,stella2018multiplex}.

The resulting logistic regression model, fitted over empirical data of word production in people with anomic aphasia, was then fed with vectorial inputs for all words in the multiplex lexical network. For the 173 words in the PNT and in the multiplex network, logistic predictions and empirical values of the probabilities for correct word naming correlated strongly (Spearman Rho $0.726$, p-value $<10^{-7}$), indicating a good consistency between regression predictions and empirical data. Through the trained regressor, for each word a probability of correct naming $p_c(w)$ was estimated. These probabilities were then used for designing aphasia-inspired attack strategies (cfr. \textit{Attack strategies} in the Methods).

\subsection{Multiplex lexical network building}

This work adopts a minimal representation of semantics and phonology in the human mental lexicon as a multiplex lexical network  \cite{stella2017multiplex,stella2018multiplex,castro2019multiplex,stella2018cohort}, where concepts are represented by nodes and are linked on two network layers by:
\begin{itemize}
\item Free associations \cite{de2018small,de2016large,kenett2011global}, indicating recall of concepts elicited by participants in a cognitive task (e.g., reading "bed" elicited the concept "sleep" $x$ times). This layer was based on the Small World of Words dataset by De Deyne and colleagues \cite{de2018small}. Associations recalled more than $x \geq 11$ times were considered, in order for the whole layer to feature as many links as the phonological one. Also for consistency with the phonological layer, this layer was treated as undirected and unweighted, as done in previous cognitive computing approaches \cite{kenett2017semantic,castro2019multiplex,stella2019viability}.
\item Phonological similarities \cite{vitevitch2008can,stella2015patterns}, linking words sounding similar (e.g. "cab" and "cat"). Sound analogies were quantified by checking that phonological IPA transcriptions of words differed for the addition/substitution/deletion of one phoneme only, as in Vitevitch's original approach \cite{vitevitch2008can}. Transcriptions were obtained from the WordData[] repository curated by WolframResearch, available through Mathematica 11.3.
\end{itemize}

The resulting multiplex lexical network was minimal in the sense that, differently from other approaches \cite{stella2018cohort,stella2019viability}, it considered only one layer of semantic relationships coupled with a phonological one. Indeed, free associations identify recalls from semantic memory \cite{de2013better} that might be related not only to semantics but also to sound similarity or visual features. Despite a small overlap with semantic features and phonological relationships, as highlighted in \cite{stella2017multiplex,stella2019modelling}, a layer similarity analysis performed in a previous work \cite{stella2018multiplex} identified free associations as structurally dissimilar from the phonological layer and more similar, for its connectivity layout, to other semantic layers like conceptual generalisations or synonyms. This quantitative result identified the layer of free associations as being mainly semantic, which justifies the main approach of this study. 

The whole multiplex lexical network included 15886 English concepts and almost 75410 conceptual associations. Table \ref{tab:statistics} reports network statistics about link density and connectivity for the individual network layers, the resulting multiplex lexical network and random null models (i.e. configuration models, where links are randomised while keeping fixed the exact degree distribution of original networks \cite{newman2012communities}). The multiplex network features a largest viable component (LVC) of 4183 words and it is fully connected, i.e. its largest connected component (LCC) includes all 15886 words in the multiplex network. Like in past approaches \cite{stella2017multiplex,stella2018multiplex,castro2019multiplex,stella2019viability}, two words are connected if a multiplex path exists between them, i.e. a network path combining free associations and phonological similarities connected the words.

\begin{figure}[ht!]
\centering
\includegraphics[width=8cm]{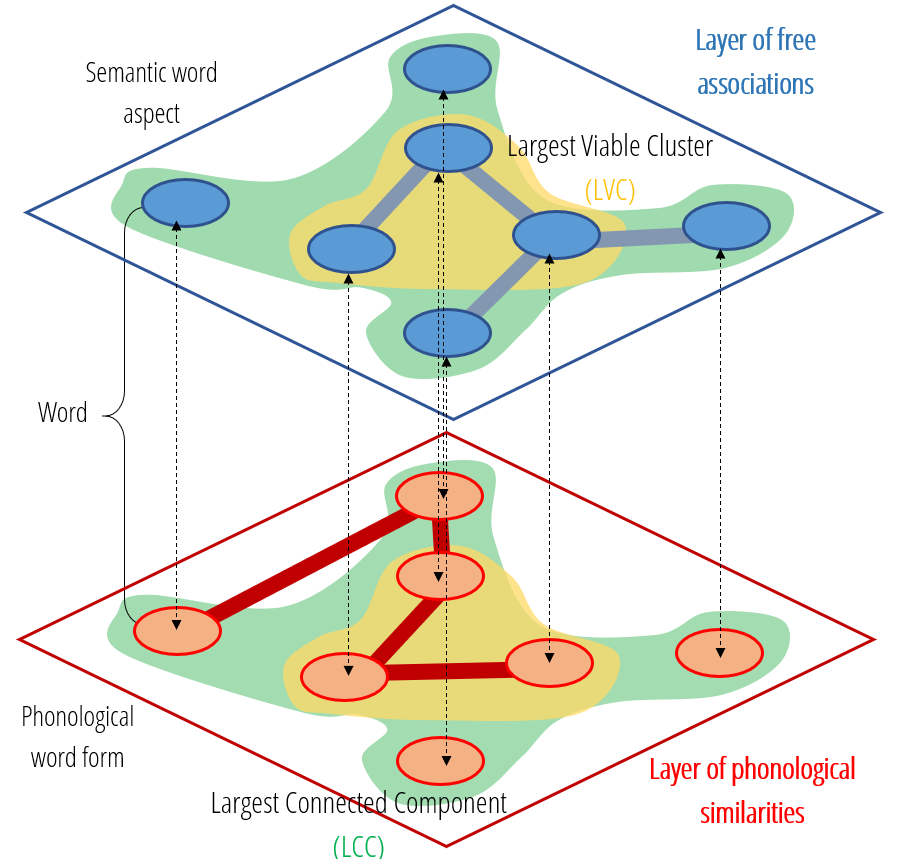}
\caption{Example visualisation of a multiplex lexical network where the LVC and LCC are highlighted. Every word is represented as two replica nodes, one representing the semantic lexical representation of a word, linked through free associations, and another one representing the phonological word form, being connected through phonological sound similarities. Jumps across layers are possible at no cost in terms of distance or transition probability.}
\label{fig:0}
\end{figure}

As reported in Figure \ref{fig:0}, viability requires connectedness on all the individual layers, i.e. viable words have a path made of semantic associations only \textit{and also} a path made of phonological similarities only both connecting them. Viable connectivity is more restrictive than connectedness when layers are combined in the multiplex structure. Instead, viability reduces to connectedness when layers are considered individually. 

\begin{table}
\caption{\label{tab:statistics}Network statistics for the layers of free associations (Free Asso.), of phonological similarities (Phon. Sim.) and the whole multiplex lexical network (Multiplex N.). LCC Size is the size of (or nodes in) the largest connected component (LCC). Mean Length stands for the average path length between any two nodes. LVC Size is the size of the large viable component  (LVC). In single-layer networks, the LVC coincides with the LCC. All layers included 15886 words. For comparison, the mean values of 10 configuration models of each layer are provided too, where the empirical degree distribution is fixed but links are randomised.}
\begin{ruledtabular}
\begin{tabular}{ccccc}
Network&Links&LCC Size&Mean Length&LVC Size\\
\hline
Free Asso.& 36383 & 11512 & 4.8 & 11512 \\
Free A. CM& 36383 & 11515 & 4.3 & 11512 \\
Phon. Sim.& 35803 & 8029 & 7.5 & 8029 \\
Phon.S. CM& 36383 & 10897 & 4.6 & 10897 \\
Multiplex N.& 75410 & 15886 & 4.6 & 4183
\end{tabular}
\end{ruledtabular}
\end{table}

The free association data from the Small World of Words project \cite{de2018small} included also the age of participants who provided a given free association. Hence, the dataset enabled the possibility to build free association networks for different age groups. Age groups were chosen in order to produce layers of free associations with the same link density as the aggregated one (for which associations produced by at least 11 individuals were selected). The resulting age groups considered weaker free associations, produced by at least 3 individuals of the following ages: (i) younger than 22 years old, (ii) between 22 and 28 years old, (iii) between 28 and 36 years old, (iv) between 36 and 50 years old and (v) older than 53 years old. Free associations map patterns of memory recall \cite{de2016large,kenett2017semantic}, which is supposed to change with ageing \cite{dubossarsky2017quantifying,wulff2019new}. In this way, considering more fine-grained layers of semantic associations filtered by age can provide temporal snapshots about how the robustness of the mental lexicon evolves with ageing.

\subsection{Viability identifies key words in the mental lexicon}

The largest viable cluster, or LVC, identified a set of words possessing different psycholinguistic features compared to others outside of it. Location equivalence tests indicated that words within the LVC: (i) had a higher median frequency, (ii) were shorter in length, (iii) were learned earlier on during language acquisition and (iv) possessed more meanings than words outside (cfr. Table \ref{tab:loceq}). These results indicate that the LVC contains words key to language processing and in this way it represents a language kernel \cite{cancho2001small}, i.e. a set of general, frequent and easier to acquire words of crucial relevance for efficient communication.

Also the empirical probability of correct picture naming $p_w$ was tested between words within and outside the LVC (see Methods). For people with anomic aphasia, pictures representing words in the LVC were easier to name compared to pictures representing concepts outside of the LVC. This difference might indicate a beneficial effect of the LVC in promoting lexical identification and word production in a clinical population but this conjecture will be further tested with attacks in the Results section.

The above results indicate that the LVC is of relevance for language representation and processing and its connectivity will therefore be used as a metric for robustness in the attack experiments.

\begin{table}[ht]
\caption{Location equivalence tests for psycholinguistic features of words outside and within the LVC (Mann-Whitney test). Length was measured in terms of letters. Polysemy identified the number of meanings of a word. A log correction was applied to frequency as performed in most psycholinguistic regression analyses.}
\label{tab:loceq}
\centering
\begin{tabular}{cccc} 
\toprule
\textbf{Psychol. Feature}	& \textbf{In LVC}	& \textbf{Out LVC} & \textbf{Test}\\
\hline
Word Length		& 5	& 7 &$7.7\cdot 10^6$, $p<10^{-7}$\\
Log Frequency		& 7.05	& 1.31 & $3.6\cdot 10^6$, $p<10^{-7}$\\
Age of Acquisition	& 7.06 yrs	& 9.57 yrs & $1.4\cdot 10^6$, $p<10^{-7}$\\
Polysemy Score		& 4	& 2 & $4.2\cdot 10^6$, $p<10^{-7}$\\
Correct Naming Prob. & 0.87	& 0.74 & $5131$, $p<10^{-7}$\\
\hline
\end{tabular}
\end{table}

\subsection{Attack strategies}

In order to mimic progressive cognitive decline in the mental lexicon, network attacks were performed progressively \cite{frantz2009robustness,iyer2013attackrobustness}, i.e. by removing increasing fractions of nodes over time from the network topology. An attacked/removed node did not contribute to connectivity and could not be traversed in order to connect other nodes. Attacked nodes did not recover over time, i.e. when more and more nodes where attacked. The number of nodes (size) in the largest connected component and in the largest viable component of the multiplex lexical network were measured over time for each and every attack strategy. Nodes were ranked according to the initial layout of the multiplex lexical network. No online, intra-layer update \cite{zhao2016robustness} was performed in this case since the robustness experiments were focused on detecting word centralities in the original, fully working networked mental lexicon.

Attack strategies were either random or targeted. In random attacks, words were selected uniformly at random and attacked/removed. Results were averaged over 100 instances of random attacks. Targeted attacks removed first nodes central/relevant for the mental lexicon. Words were ranked and removed at uniform steps (i.e., removing top 100 nodes, top 200 nodes, top 300 nodes, etc.). Centrality was estimated in terms of either psycholinguistic norms, network location or aphasia-inspired probabilities $p_c(w)$. In psycholinguistic attacks, words of either higher frequency or smaller length or lower mean age of acquisition or higher polysemy (i.e. number of different meanings) were attacked first. For network attacks, degree, PageRank, betweenness and closeness were used as metrics of word centrality. Degree counted the number of associations of a given word at a microscopic level in the conceptual structure of the mental lexicon. Phonological degree is generally also known as phonological neighbourhood density in psycholinguistics. Multidegree \cite{battiston2017new} counted all phonological and semantic associations together. In both semantic and phonological networks, words with high degree were found to be easier to learn \cite{beckage2016language,stella2019modelling} and retain \cite{siew2016spokenrecall,siew2018cognitive} in memory tasks. PageRank measured the probability for a random walker to visit a given node, indicating how words are central for random navigation of conceptual associations \cite{newman2018networks}. On the multiplex network, the definition of PageRank versatility was adopted \cite{battiston2017new,de2015ranking,bianconi2018multilayer}. In semantic networks, words with high PageRank were found to require a smaller effort in terms of memory search \cite{griffiths2007google}. Betweenness counted the number of shortest paths going through a given node either on individual layers or on the whole multiplex network. In this way, betweenness could capture how relevant concepts are as "bottle-necks" for the diffusion of spreading activation signals \cite{beckage2016language}. Both betweenness and PageRank represented meso-scale network measures  \cite{newman2018networks}, crucially depending on how clusters of concepts are associated with each other. Closeness indicated the inverse of the average number of conceptual associations connecting a word to every other concept \cite{newman2018networks}. In case of a uniform diffusion of spreading activation across all concepts, closeness would capture the potential for a concept to attract and store activation signals at a global network level \cite{siew2018cognitive}. On multiplex networks, high-closeness concepts were found to be acquired earlier during language learning \cite{stella2017multiplex,stella2019modelling} and were also identified as easier to be correctly named by people with aphasia \cite{castro2019multiplex}.

For aphasia-inspired attacks, two strategies were deployed. In a "small-to-big" attack strategy, words with a lower probability for correct naming $p_c(w)$ were attacked first, simulating a more realistic progressive decline were more difficult-to-produce words became inactive in the mental lexicon first. A less realistic reverse strategy was simulated as well. In a "big-to-small" attack, words with a higher probability for correct naming $p_c(w)$ were attacked first, simulating a worst-case scenario were the mental lexicon started failing mainly with words better processed by a population of people with anomic aphasia. This strategy is a "worst-case" scenario because it targets first words whose cognitive processing is still relatively efficient even in a clinical population with a diagnosed clinical language impairment. These two attacks represent a \textit{big data} approach to testing how the large scale structure of the mental lexicon behaves under modelled cognitive decline specifically based on clinical data.

\section{Results}

This section is divided in three different parts. The first one outlines the robustness of the global connectivity of the multiplex lexical network. The second part focuses on the specific connectedness of the largest viable component \cite{iyer2013attackrobustness,bianconi2018multilayer}. Both these network measures aim at assessing the overall robustness of the mental lexicon to progressive word failures/attacks/removals. Given the richness of the LVC \cite{baxter2016unified} in terms of words of relevance for language use, acquisition and processing (cfr. Methods), the results on its connectedness provide detailed additional information on how associations between specifically "core words" in the mental lexicon decline over progressive attacks. The third and last part tests how fragility patterns evolve over time in the mental lexicon when semantic connections age and weaken.

\subsection{Robustness of the networked mental lexicon in terms of largest connected component}

The size of the LCC identifies the number of words connected by either semantic or phonological conceptual associations. Its decline depends on the specific way of removing/attacking words. As reported in the Methods section, in all target attacks node removal was linear (i.e., removing top 100 nodes, top 200 nodes, top 300 nodes, etc.). Hence, a linear decline in LCC is considered to be the lowest damage possible to LCC size. Such linear decline is called also \textit{graceful decline} \cite{iyer2013attackrobustness,newman2018networks} and it represents evidence for robustness to the relative attack strategy. Other behaviours are possible \cite{baxter2018targeted}, e.g. non-linear fall in LCC size or phase transitions. The trend of the LCC size over the number of removed nodes is also called \textit{robustness profile} \cite{frantz2009robustness}.

\begin{figure*}[ht!]
\centering
\includegraphics[width=14cm]{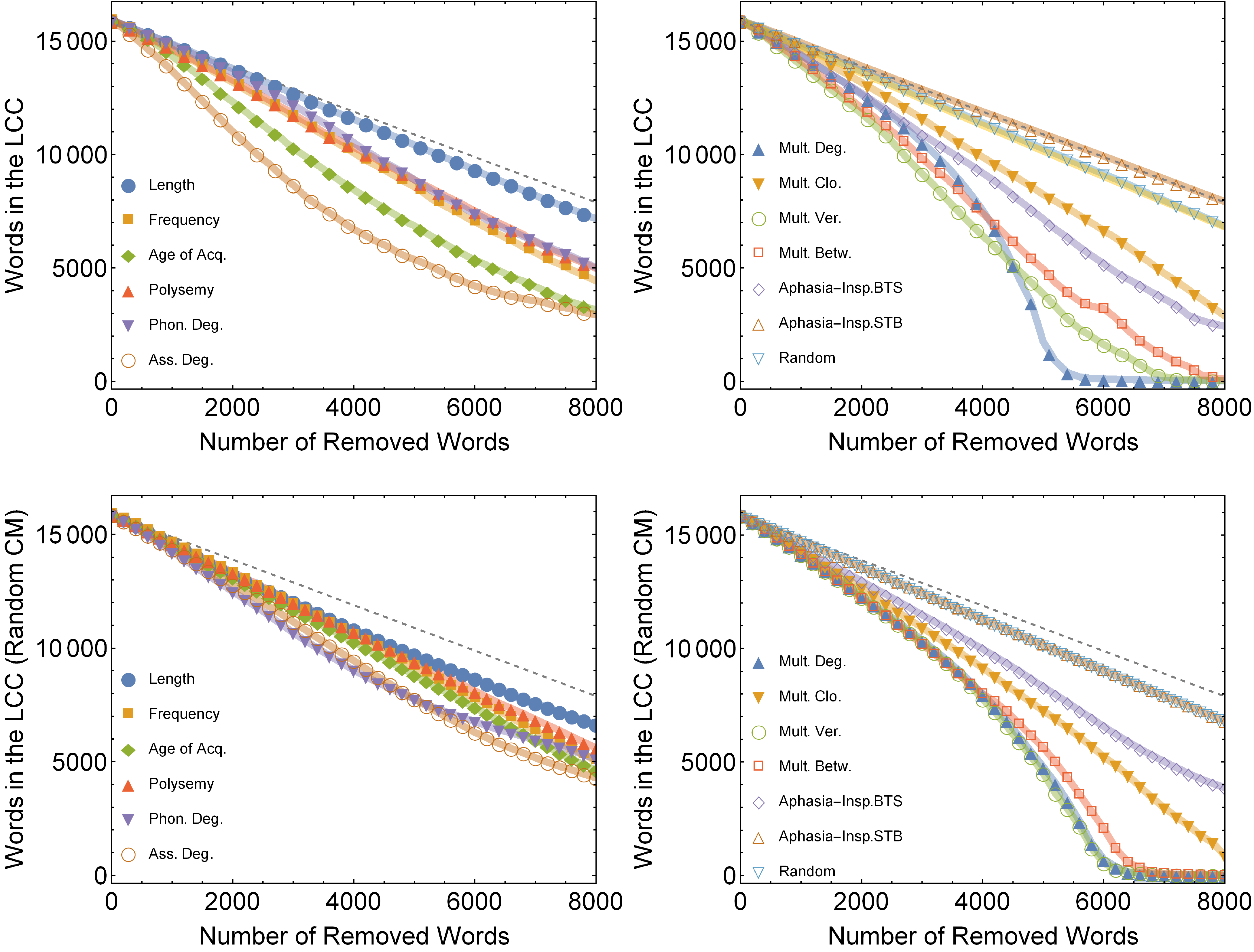}
\caption{\textbf{Top:} Robustness profiles of the largest connected component (LCC) for a multiplex lexical network representing semantic and phonological associations in the mental lexicon of an adult English speaker. \textbf{Bottom:} Robustness profiles for random configuration models with the same degree distribution of above. All results are averaged over 20 configuration models. In all subpanels, in targeted attacks, words were ranked and removed at uniform steps (i.e., removing top 100 nodes, top 200 nodes, top 300 nodes, etc.). Progressively attacking/removing nodes decreases the size of the LCC in different ways, depending on attack strategies. The dashed line represents the expected decline in LCC size due only to the decrease in total number of nodes available in the network, without considering network links.}
\label{fig:1}
\end{figure*}

Figure \ref{fig:1} reports the robustness profiles of the networked multiplex representation of the mental lexicon, adopted here, under different attacks. For visual clarity, profiles are clustered in two groups and reported in two subpanels (Fig.\ref{fig:1}, top left and right).

The networked mental lexicon is found to be resilient to random node attacks, since its LCC displays a graceful decline in size when nodes are increasingly removed (cfr. Fig.\ref{fig:1}, top right). In other words, random failures of words are not capable of crippling the connectivity of conceptual associations between words by erasing or drastically reducing the size of the largest connected component. 

A lower resilience of multiplex, networked mental lexicon is found across all psycholinguistics-driven attacks (cfr. Fig.\ref{fig:1}, top left). Removing words of shorter length, higher frequency, acquired earlier or with more meanings does not completely disrupt the LCC when $50\%$ of the concepts have been attacked. However, these attacks can significantly impair connectedness. For instance, removing up to top 50$\%$ words of lower age of acquisition reduces the LCC up to $25\%$ of its original size. Age of acquisition is found to be the most disruptive psycholinguistic strategy, indicating a structural organisation of concepts in the mental lexicon significantly relying over age of acquisition rather than on frequency or other metrics. Alternatively put, the robustness profile indicates that words acquired earlier are more relevant than words of higher frequency, shorter length or more meaning in order to connect different concepts together across semantic and phonological aspects of language. This finding is in agreement with previous small-scale studying showing the predominance of age of acquisition over frequency in driving lexical retrieval \cite{morrison1992age}.

Network-driven attacks can dismantle the networked representation of the mental lexicon significantly more than psycholinguistic strategies (cfr. Fig.\ref{fig:1}, top left and right). A catastrophic \cite{casti1979connectivity} phase transition over the size of the LCC is found in the multiplex degree-based attack strategy (Fig.\ref{fig:1}, top right). Whereas removing words with the most free associations (phonological similarities) cripples the LCC to a limited extent, analogous of psycholinguistic attacks (Fig.\ref{fig:1}, left), instead removing nodes based on the \textit{multiplex combination} of semantics and phonology leads to a disruptive attack strategy, where the LCC vanishes completely through a catastrophic phase transition after only $33\%$ of words are removed from the multiplex network. Such phase transition could not be observed in case semantic and phonological layers were observed individually and it is therefore a phenomenon emerging from the multiplex structure of the mental lexicon. The complete disruption of a largest connected component containing a large fraction of nodes when only a part of nodes is attacked indicates a lack of robustness or rather a network fragility \cite{iyer2013attackrobustness,baxter2018targeted}. In other words, the modelled representation of the mental lexicon is found to \textit{ be fragile} to local attacks mediated by multiplex degree. The same lexical structure is more resilient to attacks based on single-layer network degree, indicating the relevance of considering semantics and phonology together as a multiplex network when investigating the robustness of the mental lexicon to progressive cognitive decline. 

Other multiplex metrics like PageRank versatility and multiplex betweenness could dismantle the LCC completely like multiplex degree, though by using considerably more attacks. Single-layer PageRank, betweenness and closeness attacks reported analogous results to single-layer degrees and were not included in Fig. \ref{fig:1} for brevity. In general, in large-scale single-layer networks all these measures are strongly positively correlated, so that analogous results are expected. Interestingly, the multiplex lexical network resulted being resilient to attacks based on multiplex closeness, which were unable to dismantle the whole LCC as quickly as other multiplex attacks did. This resilience indicates that the layout of paths connecting concepts is not localised only in shortest paths, which are captured by closeness, but also in longer, more winded, paths of conceptual associations. In other words, connectivity of concepts in the mental lexicon does not rely solely on shortest paths but it is rather more de-localised, underlining the need for developing additional measures of path-based centrality beyond closeness itself \cite{stella2018distance}.

Aphasia-inspired attacks reported two distinct behaviours. Removing first words with smaller probability of correct naming $p_c(w)$ produced even less damage than random attacks up until when 50$\%$ of words were removed. This pattern represents evidence that patients with anomic aphasia do not lose the ability to name correctly words uniformly at random but rather tend to fail in finding words \textit{peripheral} in the multiplex lexical network and for language use and acquisition, extending previous small-scale results \cite{hillis2007aphasia,erdeljac2008syntactic,castro2015using,castro2019multiplex}. Removing these peripheral words leaves the mental lexicon relatively unscathed in terms of conceptual connectedness, i.e. the LCC size falls gracefully under removal of large portions of peripheral words. The overall multiplex structure of the mental lexicon is resilient to progressive failing of words as indicated by behavioural data in anomic aphasics. Assuming that language degradation in people with aphasia corresponded completely to a degraded lexical representation \cite{erdeljac2008syntactic,borge2011modeling}, ruling out functional search deficits \cite{hillis2007aphasia} on such structure and considering only increasingly more and more failing words, this result indicates that the structure of the mental lexicon itself would degrade and fall in a way as to minimally impact conceptual connectivity, a graceful decline indicating a \textit{strong resilience} to cognitive degradation in anomia. 

If the mental lexicon is robust to removing words whose lexical retrievals fails first in people with anomic aphasia and such words are peripheral in the multiplex structure of the mental lexicon, then what about robustness to removal of core words, instead? Fig.\ref{fig:1} (top right) reports the robustness profile of the mental lexicon when words easier to name for anomic aphasics are removed first, i.e. "big-to-small" attacks. The LCC does not vanish as quickly as in the case of multiplex attacks, but it is still reduced to $16\%$ of its original size when only 50$\%$ of words in the whole network are attacked/removed. Hence, these words with high $p_c(w)$ are indeed core words in the global lexicon structure. This result indicates that aphasia-inspired attacks can capture relevance of words in the mental lexicon and produce sensible disruptions of conceptual connectivity. The effects of such attacks drastically depend on the type of words targeted by them, e.g. peripheral vs core.

Resilience to random attacks but fragility to degree-based attacks is a feature displayed by scale-free networks \cite{newman2018networks}, where hubs are fundamental for connecting nodes since they detain a large fraction of links. As reported also in a previous work \cite{stella2019viability}, both the free associations and phonological layers reported here have a heavy-tailed degree distribution featuring hubs. In order to understand to what extent presence of hubs determines the robustness/fragility of the empirical multiplex network, a null model made of configuration layers is considered. Configuration models randomise links by keeping fixed the degree of every node \cite{newman2018networks}. In this case, configuration models disrupt the meaning of conceptual associations \cite{stella2018multiplex} but keep fixed the phonological neighbourhood density and the semantic richness (i.e. the degrees) of every word. 

Figure \ref{fig:1} reports the average robustness profiles estimated over a sample of 20 configuration models of the original semantic and phonological layers. Psycholinguistic norms were kept but conceptual associations were randomised. The null models display different robustness features in comparison to the empirical multiplex network.

Psycholinguistic norms are less effective in disrupting the random multiplex network, see also Fig. \ref{fig:1} (bottom left). The most drastic difference is observed over attacks based on the age of acquisition. This quantitative result indicates the presence of correlations between psycholinguistic norms and conceptual links beyond the local scale/node degree (which is preserved by configuration models), further underlining the need for developing cognitive network metrics in conjunction with psycholinguistic norms beyond the scale of word degree, richness and neighbourhood density.

The random multiplex network is still fragile to multiplex degree but more robust to single-layer degree attacks. Fig. \ref{fig:1} (bottom right) highlights a continuous phase transition in the size of the LCC when more and more nodes are attacked based on multiplex degree. However, notice that this transition happens when 1000 more words are removed in comparison to the empirical case. Hence, the empirical multiplex network is more fragile than random expectation to multiplex degree attacks. Degree, which is preserved in configuration models, is not enough for explaining this fragility by itself. Such difference indicates that words with many semantic and phonological associations are highly relevant in the empirical multiplex network not only because of their degree but also because of the overall meso-scale and global connectivity of the mental lexicon. These robustness profiles represent a strong quantitative evidence that in the mental lexicon degree correlates also with large-scale features of conceptual associations, which together enable the identification of central words and whose removal dismantles \textit{catastrophically} \cite{casti1979connectivity} (i.e. through a phase transition) connectivity among concepts.

Another difference with the empirical case is that in the null model multiplex betweenness, PageRank and multidegree give rise to an analogous phase transition whereas in the empirical network these attack strategies provided strongly different results. This represents additional evidence that heavy-tailed degree distributions are not enough for explaining the robustness and fragility patterns observed in the empirical multiplex lexical network. The empirical multiplex network is more robust than random expectation to meso-scale attacks, as captured by betweenness and Page-Rank. This indicates the presence of empirical conceptual associations in the human mental lexicon providing robustness to degradation and disrupted by the random rewiring of configuration models.

\subsection{Robustness of the networked mental lexicon in terms of largest viable cluster}


\begin{figure*}[ht!]
\centering
\includegraphics[width=16cm]{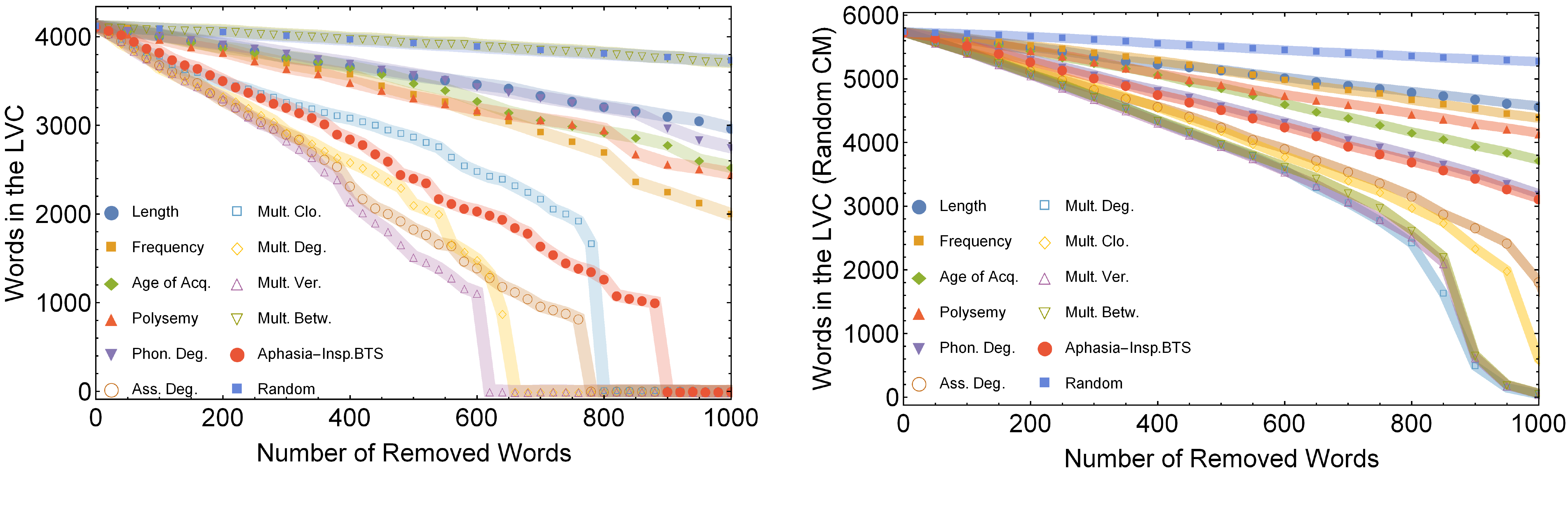}
\caption{\textbf{Left:} Robustness profiles of the largest viable cluster (LVC) for a multiplex lexical network representing semantic and phonological associations in the mental lexicon of an average adult English speaker. \textbf{Right:} Robustness profiles for random configuration models with the same degree distribution of above. All results are averaged over 20 configuration models. In all subpanels, in targeted attacks, words were ranked and removed at uniform steps (i.e., removing top 50 nodes, top 100 nodes, top 150 nodes, etc.). Progressively attacking/removing nodes decreases the size of the LVC in different ways, depending on attack strategies. Removing words of lower $p_c(w)$ resulted in minimal damage, equivalent to word length attacks, and was not reported for brevity.}
\label{fig:2}
\end{figure*}

Figure \ref{fig:2} reports the robustness profiles for the empirical LVC (left) and its randomised counterpart through configuration models (right). The empirical LVC is fragile to multiplex PageRank attacks, indicating that words in the viable cluster are versatile \cite{battiston2017new,de2015ranking,bianconi2018multilayer} and hence fundamental for transitioning between the semantic and phonological aspects of the mental lexicon through a random walk. The empirical multiplex network is fragile, to a lesser extent, also to attacks based on multiplex closeness, association degree and multidegree. Aphasia-inspired attacks targeting words with a higher probability of correct naming are able to fully dismantle the LVC by removing only 20$\%$ of its nodes. All these attacks display also a sudden discontinuity over the LVC size, reminiscent of the explosive phase transition detected during cognitive development by Stella and colleagues \cite{stella2018multiplex}. Explosive phase transitions were observed also in multidegree attacks of multiplex networks made of random graphs \cite{baxter2018targeted}. However, these discontinuous transitions were not reproduced by the configuration models investigated here (see Figure \ref{fig:2}, right), mainly because of the heavy-tailed degree distribution observed in free associations and phonological similarities.

No psycholinguistic-based attack is able to fully dismantle the LVC, indicating that the information encapsulated in conceptual associations and captured by network metrics are qualitatively different from the information identified by linguistic norms. This finding represents additional evidence that psycholinguistic measures of word relevance and network centralities provide different information and should be therefore used in a \textit{synergistic combination} in order to obtain a richer description of word relevance for language acquisition and use.

The fragility of the empirical LVC to multiplex PageRank, multidegree and multiplex closeness cannot be explained by degree itself as in the random models no discontinuous phase transition is present when only 800 or less words are attacked. This indicates that the catastrophic, sudden phase transitions observed in the empirical multiplex network are a linguistic pattern that is not a direct sequence of heavy-tailed degree distributions. In other words, the LVC being more fragile than random expectation to these kinds of attacks indicates that concepts included in the viable component are relevant to the mental lexicon not only in terms of their local degree but also in terms of meso-scale and large-scale network patterns.

Removing words of lower correct naming probability $p_c(w)$ for people with aphasia resulted in minimal damage and a graceful decline of the LVC size equivalent to word length. This indicates that the words that people with aphasia find the most difficult to name in picture naming tasks are mainly words in the periphery of semantic and phonological networks, outside the core of relevant words identified by the LVC. 

On the other hand, the LVC is fragile to removal of those words that are found easier to be named by people with anomic aphasia, i.e. attacking words with higher $p_c(w)$ first. This represents a strong quantitative indication that being within the viable cluster \textit{correlates positively with a facilitative effect} over word retrieval. 

Although this analysis cannot shed light over causal relationships, it highlights the clinical relevance of the LVC as a sub-component of the mental lexicon where having multiple semantic and phonological paths between words correlates with a \textit{shielding effect} against cognitive decline and failing lexical processing. Words in the LVC are found to be easier to be produced than words outside of the LVC, hence the fragility to the "big-to-small" aphasia-inspired attack.

\subsubsection{Persistence of catastrophic transitions in the robustness of the mental lexicon across age groups}

\begin{figure}
\centering
\includegraphics[width=7.5cm]{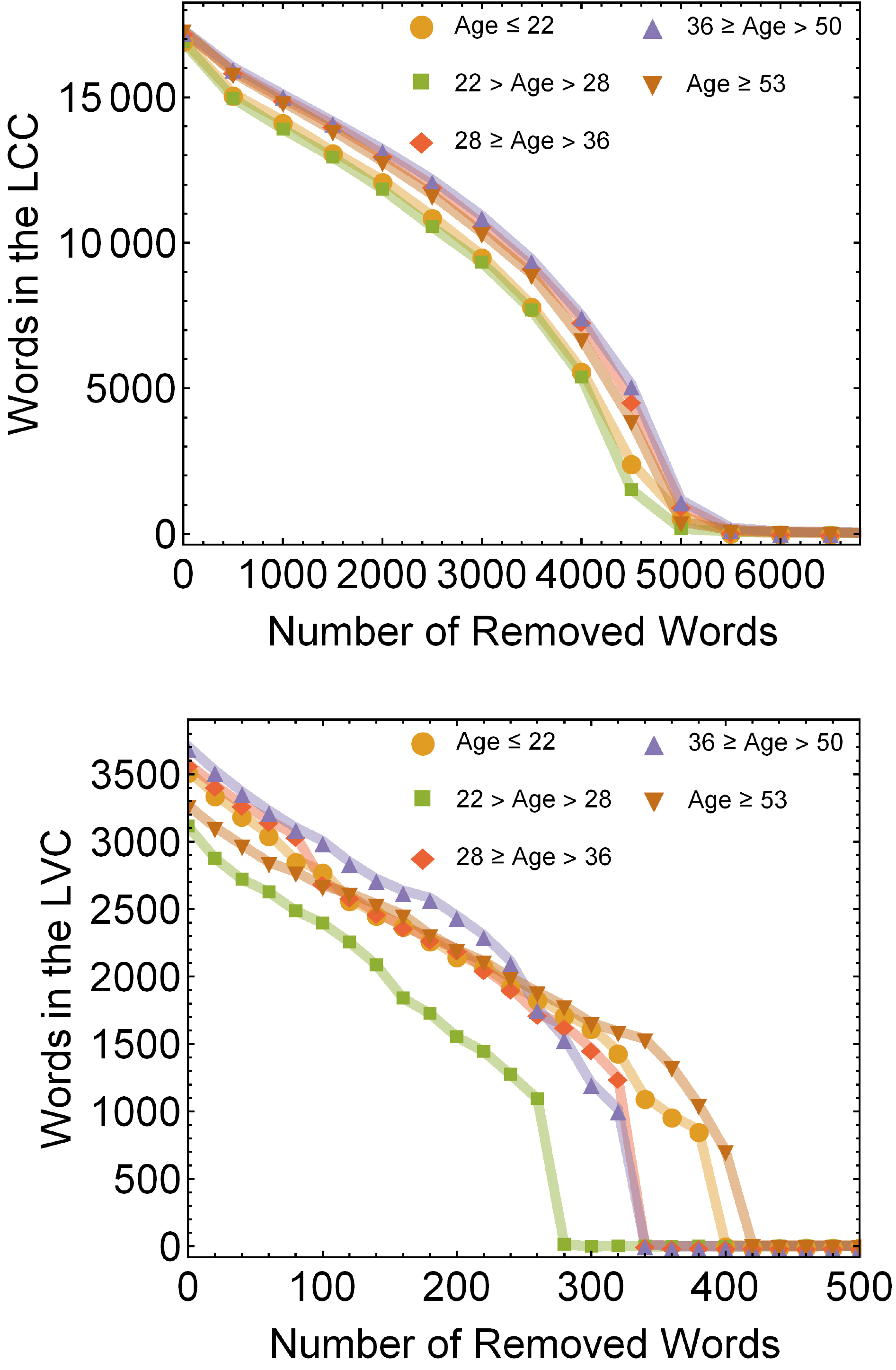}
\caption{\textbf{Top:} Robustness profiles under multidegree attacks of the largest connected component (LCC) for a multiplex lexical network representing semantic and phonological associations in the mental lexicon of an adult English speaker of a given age group. \textbf{Bottom:} Robustness profiles under multiplex PageRank for the largest viable cluster (LCC) for a multiplex lexical network representing semantic and phonological associations in the mental lexicon of an adult English speaker of a given age group.}
\label{fig:3}
\end{figure}

The mental lexicon is not a fixed cognitive system but it rather evolves over time. Free associations indicate memory recall patterns that can weaken with ageing \cite{wulff2019new}, thus leading to more rarefied or less clustered network structures over time, as reported also in previous studies \cite{dubossarsky2017quantifying}. In this way, it is important to assess the robustness of the mental lexicon across different age groups, in order to monitor potential changes due to ageing. Ageing was modelled only at the semantic level, for which different networks of free associations were used (cfr. Methods).

Figure \ref{fig:3} reports the robustness profiles of the LCC (LVC) under multidegree (multiplex PageRank) attacks for different ageing stages. These two attack strategies were selected and tested across age groups because they resulted being the most disruptive for the LCC and LVC, respectively. Despite fluctuations due to small differences in the link density of each network, all multiplex representations of the mental lexicon across age groups featured a catastrophic phase transition in the LCC size under multidegree attacks when roughly 5000 words were attacked. The attack of the LVC by means of multiplex PageRank displayed a larger variability across age groups, with the layer of young adults between ages 22 and 28 displaying the most fragile LVC. Across all age groups, the largest viable component vanished when a rather small fraction of high multiplex PageRank words were removed.

As a result, the same fragility patterns observed above for an age-aggregated mental lexicon also \textit{persist} across different age-specific multiplex lexical networks. 

\section{Discussion}

This study investigated the robustness of the mental lexicon towards progressive word failure through a network approach. The associative structure of concepts in the mental lexicon was modelled as a \textit{minimal} multiplex network combining two aspects of language: semantic relationships (through free associations \cite{de2013better,de2016large,de2018small}) and sound patterns (through phonological similarities \cite{vitevitch2008can,neergaard2019constructing}). 

The combination of these two aspects of the mental lexicon led to results absent in the individual single-layer networks. In fact, the whole networked mental lexicon exhibited a catastrophic \cite{casti1979connectivity} phase transition over node connectivity when nodes of high multiplex degree where attacked first. Attacking nodes based only on their semantic richness \cite{kenett2019can,kenett2017semantic,de2013better}
(i.e. free association degree) or their phonological neighbourhood density \cite{vitevitch2008can,siew2016spokenrecall} (i.e. phonological degree) did not exhibit any phase transition on the size of the largest connected component of concepts in the mental lexicon. Attacking words according to the sum of these degrees, i.e. their multidegree \cite{battiston2017new}, \textit{dismantled completely} connectedness when only half available concepts where attacked. Such evidence indicates a fragility in the layout of conceptual associations that can be observed only when semantics and phonology are represented in conjunction. This motivates the need for pursuing lexical representations \cite{vitevitch2018spoken,stella2018cohort,doczi2019overview,castro2019multiplex} accounting for semantic and phonological interactions in  acquiring, storing and processing language.

Considering null models preserving node degree but randomising conceptual links, the empirical networked mental lexicon was found to be more fragile than random expectation. This quantitative comparison indicates that hubs in the cognitive layout of conceptual associations are important not only because of their degree but also because of their overall organisation and connectivity with other concepts. Removing hubs efficiently disrupted connectedness across concepts. Connectedness is considered to be fundamental for lexical retrieval in the commonly adopted revised model of spreading activation \cite{bock1994language,kenett2017semantic,goldstein2017influence,siew2018cognitive,doczi2019overview}. Activation is a signal propagating over semantic and phonological conceptual associations and accumulating over target concepts that have to be identified or recalled. The absence of connectedness disrupts lexical retrieval, inhibiting language processing within the mental lexicon. In this way, the identified catastrophic phase transition is a trademark of the importance of hub concepts in the mental lexicon for guaranteeing semantic and phonological associations, which, combined, connect most concepts together and offer channels for the flow of activation spreading, word processing, identification and recall \cite{aitchison2012words}. 

Although fragile to multiplex degree-attack, the mental lexicon was considerably robust to progressive word failure driven by psycholinguistic features, e.g. attacking first high frequency or shorter words. The graceful decline \cite{iyer2013attackrobustness} observed in connectivity when attacking words based on their frequency, length, age of acquisition or number of different meanings indicates that the mental lexicon structure includes additional information that is not fully captured by individual psycholinguistic features. This result indicates the importance of investigating the mental lexicon structure not only in terms of individual word features \cite{brysbaert2017test} but also in terms of network metrics, including information on how concepts are connected and organised through conceptual associations.

Indeed, network attacks were able to dismantle the language kernel \cite{cancho2001small} represented by the largest viable cluster \cite{baxter2016unified} (LVC) more efficiently than random expectation and psycholinguistic metrics. In agreement with previous studies \cite{stella2018multiplex,stella2019viability}, the LVC was found to be richer in more commonly used words. Furthermore, previous approaches \cite{stella2018multiplex} identified the LVC as providing short connections between concepts and highlighted different access strategies to the LVC itself during the mental navigation performed in fluency tasks by people with high and low creativity levels \cite{stella2019viability}, respectively. The analysis of aphasic data \cite{roach1996philadelphia} reported here identified the LVC as richer in words whose correct naming is easier for people with a diagnosis of anomic aphasia \cite{hillis2007aphasia,laine2013anomia} in comparison to the rest of the mental lexicon. This facilitative effect in producing a word across aphasic populations, in addition to the above results, identifies the \textit{LVC as a region of cognitive relevance} for language processing within the mental lexicon structure.

Multiplex degree was effective in dismantling also the LVC but not as much as multiplex PageRank. Previous studies \cite{battiston2017new,de2015ranking,bianconi2018multilayer} reported multiplex PageRank as capable of identifying bottle-neck nodes fundamental for driving random walks across two distinct network layers. Given that words in a viable component have to be connected by paths made only of semantic links and, at the same time, by paths made only of phonological associations, it can be expected for such viable nodes, as captured by multiplex PageRank, to act as \textit{bridging concepts} for the whole mental lexicon. In this way, an interesting future research direction opened by the above results is assessing the role of the LVC for language processing of concepts far apart in the mental lexicon, through the design of specific psycholinguistic tasks.

Also the LVC was found to be resilient to attacks based on psycholinguistic norms but it displayed a higher fragility to aphasia-inspired attacks, which at the same time did not critically dismantle connectedness in the whole multiplex network. These aphasia-inspired attacks modelled cognitive decline as informed from data about correct word production and picture naming in people with anomic aphasia. The current approach represents a first-of-its-own extension of pioneering network investigations of cognitive decline \cite{borge2011modeling,borge2012topological} to the richer phenomenology of multiplex networks and psycholinguistics-driven attacks, a possibility opened by the recent availability of Big Data also in clinical populations \cite{mirman2010large,brysbaert2017test}.

Attacking first words that are less correctly named by people with anomic aphasia did not hamper either the LCC or the LVC in any drastic way, providing quantitative support to the \textit{graceful degradation hypothesis} in cognitive neuroscience \cite{hillis2007aphasia,laine2013anomia} that cognitive decline in anomia and related cognitive impairments starts preferentially by impacting linguistic skills in non-severe ways. In the current analysis, the first words that failed in aphasia-inspired attacks were indeed peripheral for the connectivity and viability of the whole, multiplex, associative structure of the mental lexicon structure, thus having an expected limited negative impact over linguistic skills. Understanding the correlations between structural degradation and linguistic skills remains an open challenge \cite{erdeljac2008syntactic,aitchison2012words,laine2013anomia,castro2015using} of relevance for future research.

Attacking first words more correctly named by people with anomic aphasia did not dismantle effectively connectedness but it provoked an explosive, discontinuous \cite{baxter2016unified,baxter2018targeted} phase transition in the LVC, effectively destroying it through a relatively small number of attacks. This difference in fragility provides evidence that words whose lexical retrieval is damaged less by anomic aphasia are mainly located within the LVC but are not as fundamental as hubs for connectedness of the whole mental lexicon. Alternatively, words correctly identified and produced by people with anomic aphasia are not mainly hub words but rather versatile nodes, as identified by multiplex PageRank and being part of the largest viable cluster. These results enable the hypothesis that the LVC \textit{facilitates the production of words important for transitioning} between phonology and semantics and not for achieving global connectedness. These versatile words could be shielded against cognitive decline. Unfortunately the data-informed approach used here enables the identification of a correlation between "correct naming", "being in the LVC" and cognitive decline, without allowing for the identification of causal links. The above results open an interesting way for studying cognitive decline in people with aphasia by assessing better the role that words in the LVC might have in hampering or promoting the re-training \cite{dell1997lexical,hillis2007aphasia,kirrie2018} of linguistic skills.

Interestingly, the above patterns are preserved even through ageing, when the multiplex structure is changed in order to account for alterations \cite{dubossarsky2017quantifying,wulff2019new} in semantic memory patterns (i.e. free associations) provided by young adults, adults and older adults. At all ageing stages, the multiplex mental lexicon is found to be fragile to multiplex degree-attacks while the LVC remains fragile to multiplex PageRank attacks. This persistence might be an indication that although memory patterns can weaken with ageing \cite{wulff2019new}, the organisation of the mental lexicon in terms of interacting phonological and semantic systems always presents both hubs guaranteeing global connectedness and viable nodes brokering spreading activation flow between semantic and phonological layers.

It is important to underline that the above results are based on specific assumptions and suffer from a few limitations. All the attacks reported here simulated progressive word failure where, once attacked, a concept did never recover \cite{borge2012topological,iyer2013attackrobustness}. Although in line with progressive disruptive clinical pathologies where language can be completely lost, like in Alzheimer's Disease \cite{aitchison2012words}, in other cognitive pathologies language skills can be recovered even after a long time since the onset of the disease, like in some instances of aphasia \cite{stefaniak2019neural}. The gathering of more specific data, linking language recovery and linguistic skills to different snapshots of the mental lexicon, could be used for designing realistic recovery probabilities after word attack/removal. The main issue in this case, common also to the aphasia-inspired attacks used here, is sample size: Data was available only for 173 words from the PNT \cite{roach1996philadelphia,mirman2010large}. Although logistic regression is expected to work correctly also within small sample sizes \cite{stella2019modelling}, the extension of the Philadelphia Naming Task to larger pools of words would be highly advantageous for obtaining more realistic and detailed estimates of word failures in people with cognitive impairments. 

Another limitation of the current study is the assumption that the mental lexicon structure is fixed over time and is the same across different people. Although the first downside was partially addressed by considering time-dependent free associations, at the best of the author's knowledge there is no large-scale dataset about how phonological relationships evolve over time yet. The so-called Transmission Deficit Hypothesis \cite{wulff2019new} indicates that older adults can incur into semantic activation of words that do not correspond to a phonological activation, thus inhibiting word production and leaving with a vague awareness of knowing a word without being able to produce it, a phenomenon known also as tip-of-the-tongue \cite{tiptongue1966}. In this way, ageing might have a more predominant effect over semantic rather than on phonological links, thus justifying the approximation of considering only weakened semantic associations age groups. Nonetheless, within the robustness framework outlined in this work, novel datasets investigating the structural decline of phonological and semantic associations, combined, would prove extremely valuable for better understanding failed lexical retrieval. In order to address also linguistic variability, individual networks of free and phonological similarities might be built through appropriate data collection \cite{kenett2014investigating} and investigated within the same robustness framework summarised here.

Notice also that the current analysis is limited to investigating only structural changes, and not functional nor dynamical ones, in the mental lexicon. This approximation does not keep into account cognitive search strategies \cite{hillis2007aphasia,aitchison2012words} over such structure, that might also degrade along with cognitive impairments \cite{erdeljac2008syntactic,dell1997lexical,martin1998lexical} or ageing \cite{wulff2019new}. The debate between structure and dynamics represents a relatively open field of investigation in psycholinguistics, where the most commonly accepted hypothesis is that search strategies and mental lexicon structure are strongly inter-dependent \cite{aitchison2012words}. Recent investigations like fluency performance in people with different creativity levels \cite{stella2019viability} have indicated that structural multiplex measures in synergy with psycholinguistic data are powerful enough for highlighting different search strategies on the mental lexicon, detecting differences in the way people with high and low creativity levels explore and search concepts in the animal category task. Similarly, other approaches based on multiplex structure have been successful in predicting picture naming performance of people with aphasia \cite{castro2019multiplex}. Through a different network construction, structural network metrics were powerful in predicting also cognitive degratation due to Alzheimer's disease \cite{zemla2019analyzing,zemla2018estimating}. All these approaches indicate that investigating the network structure of the mental lexicon can be considered a first important step for understanding language processing, providing the foundations for analysing other important elements of the lexicon, such as search strategies, through additional research.

The framework of network robustness outlined here opens new ways for transdisciplinary investigations of the interplay between language processing and the degrading multi-layered structure of the mental lexicon. Such ambitious aim can be achieved only through a synergy of psycholinguistic word features, cognitive science theory, clinical datasets and access to large-scale, multiplex representations of conceptual associations. Quantitative models exploiting such synergy represent an innovative way for developing data-informed, potentially individually developed models of cognitive decline and language restoration \cite{kirrie2018} in clinical patients.

\section{Acknowledgements}

The author acknowledges Dr. Nichol Castro, University of Washington, for insightful discussion in the early stages of this work.

\bibliography{aipsamp}


\end{document}